# On the Large Nc Expansion in Quantum Chromodynamics

William A. Bardeen
Theoretical Physics Department
Fermilab, MS 106, P.O. Box 500
Batavia, Illinois 60510
bardeen@fnal.gov

Abstract

I discuss methods based on the large Nc expansion to study nonperturbative aspects of quantum chromodynamics, the theory of the strong force. I apply these methods to the analysis of weak decay processes and the nonperturbative computation of the weak matrix elements needed for a complete evaluation of these decays in the Standard Model of elementary particle physics.

- Introduction.

Field theories are frequently studied through a perturbative expansion in the interaction strength or coupling constant. In many cases, a nonperturbative analysis is required to apply these theories in physical situations. The large N expansion is a nonperturbative method of analysis that makes use of particular limits for parameters unrelated to the coupling constant. For example, O(N) spin systems where N is the number of spin components can be studied by mean field methods which become exact in the large N limit. Applications of perturbative QCD, such as the cross-sections for high p_t jets, are greatly simplified using color-ordered amplitudes and a large Nc expansion where Nc is the number of colors, the quantum number associated with the QCD gauge dynamics. The large Nc expansion can also be used to study nonperturbative aspects of quantum chromodynamics with applications to the structure of chiral symmetry breaking and hadronic string theory. More recently, it has been established that there exists a duality between the large N limit of SU(N) supersymmetric Yang-Mills theory and classical supergravity in a higher dimensional space-time. In this talk, I will discuss the nature of the large Nc expansion in QCD and its application to the computation of weak decay amplitudes.

- Large Nc expansion in QCD.

The large Nc expansion defines a nonperturbative reordering of the QCD coupling constant expansion. Each Feynman diagram can be classified by its dependence on the strong coupling constant, $\alpha_{strong}$, and the number of colors Nc. In 1974, 't Hooft [1] argued that the structure of the theory simplified in the limit, $Nc \rightarrow \infty$, $\alpha_{strong} \rightarrow 0$, $\alpha_{strong} * Nc\ fixed$. The theory is summed to all orders in the rescaled coupling constant, $\alpha_{strong} * Nc$, with corrections being formally suppressed by powers of 1/Nc. At leading order in this expansion, the gluons are constrained to be in planar Feynman diagrams. Quarks lines form boundaries of the planar surfaces formed by the gluons. In this sense, the structure of large Nc QCD is similar to an open string theory with quarks attached to the ends of the string.

If we assume that large Nc QCD is confining, then the physical states are expected to consist of stringlike glueballs and mesons formed as quark–antiquark boundstates. The physical states are towers of meson resonances with increasing mass and spin. The lowest order meson scattering amplitude has the structure of a gluonic disk with meson bound-states attached to the edge of the disk. Since the color wavefunction of a meson bound-state is $O(1/\sqrt{Nc})$, the leading order meson amplitudes with more than two mesons bound-states attached to the edge of the disk are suppressed by a factor of $(1/\div Nc)$ for each additional meson. This simple power counting implies that all mesons are stable at leading order and that the meson S-matrix is that of a trivial free meson theory. We may still consider the leading contribution to processes at any order in the large Nc expansion. A meson scattering amplitude at leading nontrivial order involves only meson tree amplitudes with simple poles at positions of the meson bound-states. Higher order diagrams involve either additional quark loops which are suppressed by a power of $(1/Nc)$ or nonplanar diagrams which are suppressed by powers of $(1/Nc^2)$. The 1/Nc expansion is a topological diagram expansion with the complexity increasing at higher order. In two space-time dimensions, large Nc QCD has an exact solution while only its structure can be inferred in higher dimensions.

• Weak decay amplitudes.

Nonleptonic weak decays are mediated by the virtual exchange of W and Z bosons. These processes occur at a short distance scale, and their effects can be expressed in terms of effective local interactions between quark currents and densities at low energy [2]. The weak decay amplitudes are written as an expansion in terms of short distance Wilson coefficients, $C_i(\mu)$, and sets of local quark operators, $Q_i(\mu)$,

$$A(K \to \pi\pi) = \frac{G_F}{\sqrt{2}} V_{CKM} \sum_i C_i(\mu) \langle K | Q_i(\mu) | \pi\pi \rangle \tag{1}$$

where a normalization scale, $\mu$, is introduced to separate the short and long distance physics contributions. Details of the short distance processes and perturbative QCD dynamics are used to compute the Wilson coefficients. Operator matrix elements are sensitive to long distance physics including the mechanisms of quark confinement and the formation of hadronic bound-states that can not be computed using perturbative QCD. The systematic computation of relevant Wilson coefficient functions for weak decays at two loops, (NLO), have been made by several groups [3]. The coefficient function may be expressed as

$$C_i(\mu) = Z_i(\mu) + i * Y_i(\mu), \tag{2}$$

where the coefficient, $Y_i(\mu)$, incorporates the short distance effects of CP violation.

Operator matrix elements are more difficult to analyze as perturbative methods can not be used. Several strategies have been employed to compute the matrix elements of the quark operators appearing in the expansion of the weak Hamiltonian. All of the operators being studied have the form of products of quark currents and densities which themselves are color-singlet quark bilinears. The simplest approximation invokes factorization where it is assumed the quark currents and densities independently couple to the hadrons in the initial and final states. The factorization approximation has inherent limitations as it fails to reproduce the scale dependence of the Wilson coefficents. A second method employs numerical computations in the lattice formulation of QCD. The lattice method is constrained by the size of the numerical effort required and by problems associated with the chiral structure of lattice QCD. There has be recent progress in the direct computation of weak matrix elements using the domain wall formulation of chiral fermions on the lattice [4]. Another nonperturbative method for calculating the weak operator matrix elements invokes the large Nc expansion of QCD as discussed in the following section.

- Large Nc computation of weak matrix elements.

The large Nc analysis focuses on the computation of weak matrix elements of the four-quark operators, $Q_i$, generated by the renormalization group expansion.

$$\langle Q_i \rangle = \langle (\overline{\Psi}\Gamma_i\Psi)(\overline{\Psi}\Gamma_i\Psi) \rangle \tag{3}$$

The analysis of the large Nc expansion of QCD relies on the toplogical structure of the quark diagrams and large Nc power counting to determine the various contributions to the weak matrix elements. At the leading order of the large Nc expansion, the matrix elements are factorized

$$\langle Q_i \rangle_F = \langle (\overline{\Psi}\Gamma_i\Psi) \rangle \langle (\overline{\Psi}\Gamma_i\Psi) \rangle. \tag{4}$$

We recall that the leading order quark amplitudes have the structure of a disk with the quark line forming the boundary of the disk with planar gluons filling in the surface of the disk. Since the quarks have Nc colors, the factorized quark densities must each be associated with separate disks and, therefore, two powers of Nc are generated, one for each disk. This factorized contribution is the leading order contribution (LO). Gluonic interactions between the two disks are nonplanar and therefore suppressed by at least two powers of Nc.

At next-leading-order (NLO) in the large Nc expansion, O(1/Nc), there are two possible contributions. An internal quark loop can be added to either disk of the leading order calculation giving a suppression of at least 1/Nc. Since this insertion does not involve the second disk it contributes only to the factorized matrix element. This correction represents a meson loop correction to the matrix element of the quark current or density. It will be assumed that we are able to measure these current matrix elements which include all order corrections phenomenologically.

The second contribution at NLO is nonfactorized. It arises when both quark bilinear operators are associated with a single disk. Because only one quark loop is involved instead of two, the amplitude is suppressed by a power of 1/Nc relative to the LO factorized contribution. At this order in the large Nc expansion, the two quark currents and the various meson bound-states are associated with the single quark line forming the boundary of the single disk. Hence, this amplitude is represented by a tree-level meson amplitude having only poles at the positions of the infinite tower of meson bound-states.

We can write this amplitude as the momentum integral of a meson amplitude with two external bilinear currents,

$$\langle Q_i \rangle_{NF} = \int dk \ A_{\Gamma_i \Gamma_i}(k, -k, p_1 ... p_N). \tag{5}$$

Knowledge of meson tree amplitudes may be used to compute integrand. At low momentum, chiral Lagrangian description is an exact representation of QCD dynamics for matrix elements involving low energy meson states. At high momentum, we can use large Nc version of the QCD renormalization group equation to compute integrand in terms of perturbative coefficient functions and factorized matrix elements. The entire integral is obtained by interpolating between the nonperturbative low energy approximation and the perturbative high energy amplitude. Calculations based on this duality have achieved some success in explaining the structure of weak decay amplitudes including the octet enhancement observed in $K \to 2\pi$ decays and the CP violation seen the $\varepsilon'/\varepsilon$ measurement [5]. The precision of these methods depend on three ingredients:

- the phenomenological determination of the long distance meson amplitude,
- the order of the perturbative short distance calculation, the $\alpha_{strong}$ expansion,
- the accuracy of the interpolation between the long and short distance approximations.

The amplitude calculation requires the evaluation of the momentum integral in Eq.(5). At high momentum (short distance), $k \to \infty$, $p_1...p_N - small$, the operator product expansion (PQCD) takes the form,

$$A_{\Gamma\Gamma}(k,-k;p_1,...,p_N) \to C_{\Gamma\Gamma;Q}(k,\mu) \langle Q(\mu) \rangle_{LO} \qquad (6)$$

where the coefficient function, C, is O(1/Nc) and the operator matrix element is factorized. Note that all weak mixing processes are O(1/Nc). Using the renormalization group and the anomalous dimension matrices of perturbative QCD, we compute evolution of the NLO weak matrix elements,

$$\mu \partial_\mu \langle Q_i(\mu) \rangle_{NLO} = -\gamma_{ij}(\alpha(\mu)) \langle Q_j(\mu) \rangle_{LO} \qquad (7)$$

where the LO matrix element is factorized. The integrated form at NLO is

$$\langle Q_i \rangle_{NLO} = -\int d\mu^2/2\mu^2 \; \gamma_{ij}(\alpha(\mu)) \langle Q_j(\mu) \rangle_{LO}. \qquad (8)$$

The long distance contributions are described by tree-level meson amplitudes which involve nonperturbative parameters which are determined phenomenologically. A number of approximations to the low energy meson amplitudes has been used in the study of weak decay amplitudes. They include chiral Lagrangians, chiral Lagrangians + vector mesons, chiral quark models, and extended Nambu–Jona-Lasinio models. The chiral Lagrangian is an exact description of QCD in the light quark limit and for low momenta. The various approximations given above are different attempts to parameterize the physical low energy dynamics. The models require the input of masses and coupling constants from data other than the weak decay processes. Further efforts are needed to improve extrapolation to higher momentum scales to more precisely match the perturbative QCD calculations of the high momentum part of the two-current correlation function. These efforts may involve adding additional mesons (vector, axial-vector, scalar, tensor, …) or higher dimension operators in the effective field theory (O($p^4$), O($p^6$), …).

In addition to the precise computation of the above tree-level meson amplitudes, matching conditions are also required to connect the momentum integral of the two-current correlation function with the standard perturbative analysis of the weak matrix elements. At NLO in the large Nc expansion, the weak decay amplitudes are divergent at high energy. The standard short distance analysis uses dimensional regularization, (NDR, HV), to regularize the quark amplitudes and define normalization scales for the coefficient functions and operator matrix elements. Both the chiral Lagrangian calculation and the short distance quark calculation may be regularized by a cutoff on momentum flowing through color singlet currents.

A consistent treatment of the short distance contributions matches the different regularization schemes by computing quark matrix elements of weak operators using both dimensional regularization and momentum cutoff methods. To isolate the purely short distance effects, an infrared regularization is introduced,

$$\int dk \to \int dk \ k^2/(k^2 - M^2) \ , \ M = IR \ cutoff . \tag{9}$$

We may now compute the perturbative matrix elements of the various quark operators using the different UV cutoff schemes. The operator matrix elements will have a finite correction factor at one loop,

$$\langle Q_i^{NDR,HV} \rangle = \langle Q_i^{mom} \rangle - w_{ij}^{NDR,HV} \left( \frac{\alpha}{4\pi} \right) \langle Q_j^{mom} \rangle \tag{10}$$

where $w_{ij}$ is rotation matrix between dimensional regularization basis and momentum cutoff basis.

This rotation matrix has been computed for all ten four-fermion quark operators used to expand the weak Hamiltonian for both NDR and HV reqularization schemes [6]. The effective coefficient functions to use with momentum subtracted matrix elements can now constructed,

$$C_i^{mom} = C_i^{NDR,HV} - C_k^{NDR,HV} w_{ki}^{NDR,HV} \left( \frac{\alpha}{4\pi} \right) \tag{11}$$

An example of effect of this shift on coefficient functions can be evaluated. Our numbers are based on the perturbative QCD analysis of Bosch et al (199) [7]. The normalization scale is $\mu = 1.3 \ Gev$ and $\Lambda_{QCD} = 0.340$. The result for certain coefficient functions are given below

| CF | NDR | HV | $mom_{NDR}$ | $mom_{HV}$ |
|---|---|---|---|---|
| Z1 | -0.425 | -0.521 | -0.669 | -0.687 |
| Z2 | 1.244 | 1.320 | 1.371 | 1.394 |
| Y3 | 0.030 | 0.034 | 0.041 | 0.041 |
| Y4 | -0.059 | -0.061 | -0.063 | -0.064 |
| Y5 | 0.005 | 0.016 | 0.013 | 0.011 |
| Y6 | -0.092 | -0.083 | -0.091 | -0.090 |

There is an enhancement of the Z1 and Z2 coefficients in the momentum basis over the values in either the NDR or HV schemes. The larger values of Z1 and Z2 will tend imply a larger octet enhancement and 27 supression factors than the conventional treatment. Of course, the important point of this aspect of the analysis is that we now have all the elements to make a consistent use of the 1/Nc expansion for computing the weak matrix elements:

- the standard renormalization group analysis of the weak coefficient functions is used to compute the short distance contributions and evolve the operators to low energy scales,
- a consistent matching between the dimensional regularization schemes and the momentum cutoff is used in the NLO analysis of the nonfactorized weak matrix elements,
- chiral Lagrangians or other effective field theories are used to describe the long distance contributions to the weak matrix elements and consistently matched to the short distance contributions.

The complete calculation has yet to be fully integrated. There are still some issues regarding matrix elements involving scalar and pseudoscalar densities that can affect the evaluation of the Q_6 and Q_8 operator matrix elements needed for $\varepsilon'/\varepsilon$ analysis.

- Conclusions.

Precision tests of the Standard Model require knowledge of nonperturbative aspects of quantum chromodynamics, the strong dynamics. The Large Nc expansion combined with phenomenological knowledge of certain meson amplitudes provides one avenue for systematic estimates. The analysis outlined in this talk provides the elements for this systematic analysis of the weak decay matrix elements through next-leading-order However it is fundamentally limited by the ability to compute yet higher order terms in the large Nc expansion.

Direct numerical computation of weak matrix elements using lattice formulations of QCD provides another avenue. Recent developments related to the chiral symmetry structure of lattice QCD may lead to more realistic calculations of the necessary matrix elements.

The final picture remains unclear. As yet there is no clear violation of the Standard Model to be inferred by weak decay processes but the large value of $\varepsilon'/\varepsilon$ and the large value of the $\Delta I = 1/2$ amplitude for $K \rightarrow 2\pi$ may yet challenge its validity.

- Acknowledgements.

I thank the Alexander von Humboldt Foundation for its hospitality and support. This research is supported by the Department of Energy under contract DE-AC02-76CHO3000.

- References.